\documentclass[a4paper,10pt,twoside]{cpc-hepnp}

\usepackage{multicol}
\usepackage{graphicx}
\usepackage{booktabs}
\usepackage{amssymb,bm,mathrsfs,bbm,amscd}
\usepackage[tbtags]{amsmath}
\usepackage{lastpage}
\usepackage{mathrsfs}

\newcommand{\piz}{\pi^{0}}
\newcommand{\pip}{\pi^{+}}
\newcommand{\pim}{\pi^{-}}

\newcommand{\jpsi}{J/\psi}

\begin{document}

\fancyhead[co]{\footnotesize Wang Yadi et al: Background subtraction using probabilistic event weights}

\footnotetext[0]{Received xxxxx}

\title{Background subtraction using probabilistic event weights
}

\author{%
      Yadi Wang$^{1,2,3}$\email{wangyd@lnf.infn.it}%
\quad Beijiang Liu$^{2}$
\quad Xiaoyan Shen$^{2}$
\quad Ziping Zhang$^{1}$
}
\maketitle

\address{%
$^1$ Department of Modern Physics, University of Science and Technology of China, Hefei 230026, China\\
$^2$ Institute of High Energy Physics, Chinese Academy of Sciences, Beijing 100049, China\\
$^3$ Istituto Nazionale di Fisica Nucleare, Frascati (Rome), Italy.
}

\begin{abstract}
Background treatment is crucial to extract physics from precision experiments. In this paper, we introduce a novel method to assign each event a signal probability. This could then be used to weight the event's contribution to the likelihood
during fitting. To illustrate the effect of this method, we test it with MC samples. The consistence between the constructed background and the background from MC truth shows that the background subtraction method with probabilistic event weights is feasible in partial wave analysis at BES~III.
\end{abstract}

\begin{keyword}
background treatment, sideband, probabilistic event weights, partial wave analysis
\end{keyword}

\begin{pacs}
02.50.Cw, 13.25.Ft
\end{pacs}

\begin{multicols}{2}

\section{Introduction}
A common situation in many experiments we often face is the presence of a non-interfering background that can not be cleanly separated from the desired signal in the event classification by any selection criteria. The remained background must be considered in the subsequent analysis. Partial wave analysis (PWA) is a powerful tool to study the hadron spectroscopy, that allows one to extract the resonance parameters, spin-parities and decay properties with high sensitivity and accuracy. Background treatment is a crucial issue to the maximum likelihood PWA fit. In this paper, we investigate the feasibility of applying a novel method for background subtraction using probabilistic event weights~\cite{q-factor} in BESIII physics analysis.

\section{Background treatment}

A general problem of the Maximum Likelihood method is its treatment of background. As the Maximum Likelihood method is based on single events and not on density distributions like for example the $\chi^2$ method, it is not possible to simply subtract the background events from the data. Typically, there are two approaches of background treatment in the fit.

In the case where the distribution of the background events can be parameterized, the probability density function (PDF) of a selected event $x_i$ can be defined by signal PDF and background PDF.
\begin{eqnarray}
P(x_{i}) = f_{sig}\times P_{S}(x_{i}) + (1-f_{sig})\times P_{B}(x_{i}),
\end{eqnarray}
where $P_{S}(x_{i})$ and $P_{B}(x_{i})$ are the p.d.f.s for the signal and background distributions and $f_{sig}$ is the signal fraction of the data sample. The advantage of this approach is the background level (1-$f_{sig}$) and the background shape ($P_B$) can be obtained from a fit. However, the PDF of signal and background must be explicitly defined, which requires not only a model predictions but also an efficiency as a function of measurement.

If the backgrounds can not be parameterized, the background contribution can be taken into account event by event by
rescaling the likelihood function with likelihood values of background events.
\begin{eqnarray}
L^{\prime} = L_S( data )/L_S( bg ).
\end{eqnarray}
Thus,
\begin{eqnarray}
\ln L^{\prime} = \ln L_{S}( data ) - \frac {N^{\prime}_{bg}} {N_{bg}} \times \ln L_{S}(bg).
\end{eqnarray}

The likelihood $L_{S}( data )$ and $L_{S}( bg )$ defined with the signal PDF are evaluated for data and background events. Thereby it is assumed that a sample of $N_{bg}$ selected events is representative for the $N^{\prime}_{bg}$ real background events in the selected data set. $N^{\prime}_{bg}$ as the absolute number of background events in the data set cannot be determined by the fit as $L_{S}( bg )$ changes the probability density of $L^{\prime}$ and therefore has to be determined by other means.

In PWA at BES~III, background contribution is typically handled using the background subtraction method. The probability to observe the event characterized by the measurement $\xi$ is:
\begin{eqnarray}
P(\xi)=\frac{\omega(\xi)\epsilon(\xi)}{\int d\xi\omega(\xi)\epsilon(\xi)},
\end{eqnarray}
where $\epsilon(\xi)$ is the detection efficiency and $\omega(\xi)\equiv\frac{d\sigma}{d\Phi}$ is the total differential cross section.
The likelihood function is constructed by:
\begin{eqnarray}
L=\prod\limits_{i=1}^{N} P(\xi_{i})=\prod\limits_{i=1}^{N}\frac{\omega(\xi_{i})\epsilon(\xi_{i})}{\int d\xi\omega(\xi_{i})\epsilon(\xi_{i})}.
\end{eqnarray}
The optimal solution is got by minimizing the $-\ln L$,
\begin{eqnarray}
\ln L = \sum\limits_{i=1}^{N} \ln \frac{\omega(\xi_{i})}{\int d\xi\omega(\xi_{i})\epsilon(\xi_{i})} + \sum\limits_{i=1}^{N} \ln\epsilon(\xi_{i}).
\end{eqnarray}
For a given data set,the second term is a constant and
has no impact on the determination of the parameters of the amplitudes or on the relative changes of $\ln L$ values. So,
for the fitting, $\ln L$ defined as:
\begin{eqnarray}
\ln L = \sum\limits_{i=1}^{N} \ln \frac{\omega(\xi_{i})}{\int d\xi\omega(\xi_{i})\epsilon(\xi_{i})}
\label{eq_lnlikelyhood}
\end{eqnarray}
is used.
The phase space integral has to be approximated by Monte
Carlo events which by definition including the efficiency:
\begin{eqnarray}
\int d\xi\omega(\xi_{i})\epsilon(\xi_{i})\approx\frac{1}{N_{gen}}\sum\limits_{i=1}^{N_{acc}}\omega(\xi_{i})
\end{eqnarray}
$N_{gen}$ ($N_{acc}$) is the number of generated (accepted) MC events.
In this approach, the efficiency function of measurement is no longer needed in the likelihood function.
The background contributions estimated from the background sample are included in the fit with the opposite sign of log likelihood
compared to signal.

\section{Sideband subtraction}

In many analyses, the background can be estimated by Monte Carlo (MC) simulation. However, in the energy region of non-perturbative QCD, there is no inclusive MC generator which is reliable to replicate the background. Methods of Data-driven background estimation are desired. A typical approach is to use the extrapolation of the events in the sideband region to estimate the background in signal region.


In the following discussion, we take the $J/\psi\to\omega\pip\pim$ as a signal channel to illustrate the method. We generate $2\times10^{6}$ $J/\psi\to\omega\pip\pim$ signal events according to a set of toy partial wave amplitudes. A MC sample of $2\times10^8$ inclusive $J/\psi$ decay events (the signal $J/\psi\to\omega\pip\pim$ has been removed) is used to represent the backgrounds. In the background sample, the major background channels are $\pip\pim\pip\pim\piz$, $\pi\rho f_{2}(1270)$, $a_{2}(1320)\rho$ and other non-$\omega$ backgrounds. After event selections, there are still non-$\omega$ events which cannot be removed. Fig.~\ref{mw} shows the invariant mass distribution of $\pi^+\pi^-\pi^0$. The ratio of signal events and background events in our "combined data sample" is about 1:1 after event selections. We define the $\omega$ signal region as $|M(\pip\pim\piz)-M(\omega)|<0.03$~GeV$/c^{2}$, denoted as middle hashed histogram in Fig.~\ref{mw}. The sideband regions are defined as $0.09<|M(\pip\pim\piz)-M(\omega)|<0.15$~GeV$/c^{2}$, denoted as left and right hashed histograms in Fig.~\ref{mw}. The number of non-$\omega$ backgrounds can be estimated from the extrapolation of $\omega$ sidebands. A one-dimensional fit on mass spectrum of $\pip\pim\piz$ with $\omega$ is performed with Breit-Wigner (BW) ($\frac{1}{(x-m)^{2}+\sigma^{2}/4}$) convolved with Gaussian resolution ($exp(-\frac{1}{2}(\frac{x}{s})^{2})$) plusing a second order polynomial. The fit yields the number of background is $58337.5\pm241.5$, which is consist with the number of real backgrounds tagged by MC truth $59065$.


Background estimation is one of the key issues in partial wave analysis, because the knowledge of the background distribution in the phase space is needed. For the 3-body decay of $J/\psi\to\omega\pip\pim$, if the rotation around the beam axis is ignored, the phase space has 4 dimensions: the $M(\pip\pim)$ ($\pip$ and $\pim$ from $\jpsi$), the polar angle of $\omega$ in $\jpsi$ rest system, the polar angle and azimuthal angle of $\pip$ in $\pip\pim$ helicity frame. Fig.~\ref{weight_DIY_wsb} shows the comparisons of the distributions estimated from sideband events and the real backgrounds. Table~\ref{tab_side} lists the $\chi^{2}$/nbins of the comparisons. The results indicate that the background estimation with the extrapolation on 1-Dimention $M(\pip\pim\piz)$ cannot manifest the background behavior in multi-dimensional phase space. The different kinematics of the signal region and background region could also cause the deviation in background estimation.


\begin{center}
\includegraphics[width=6cm,height=5cm]{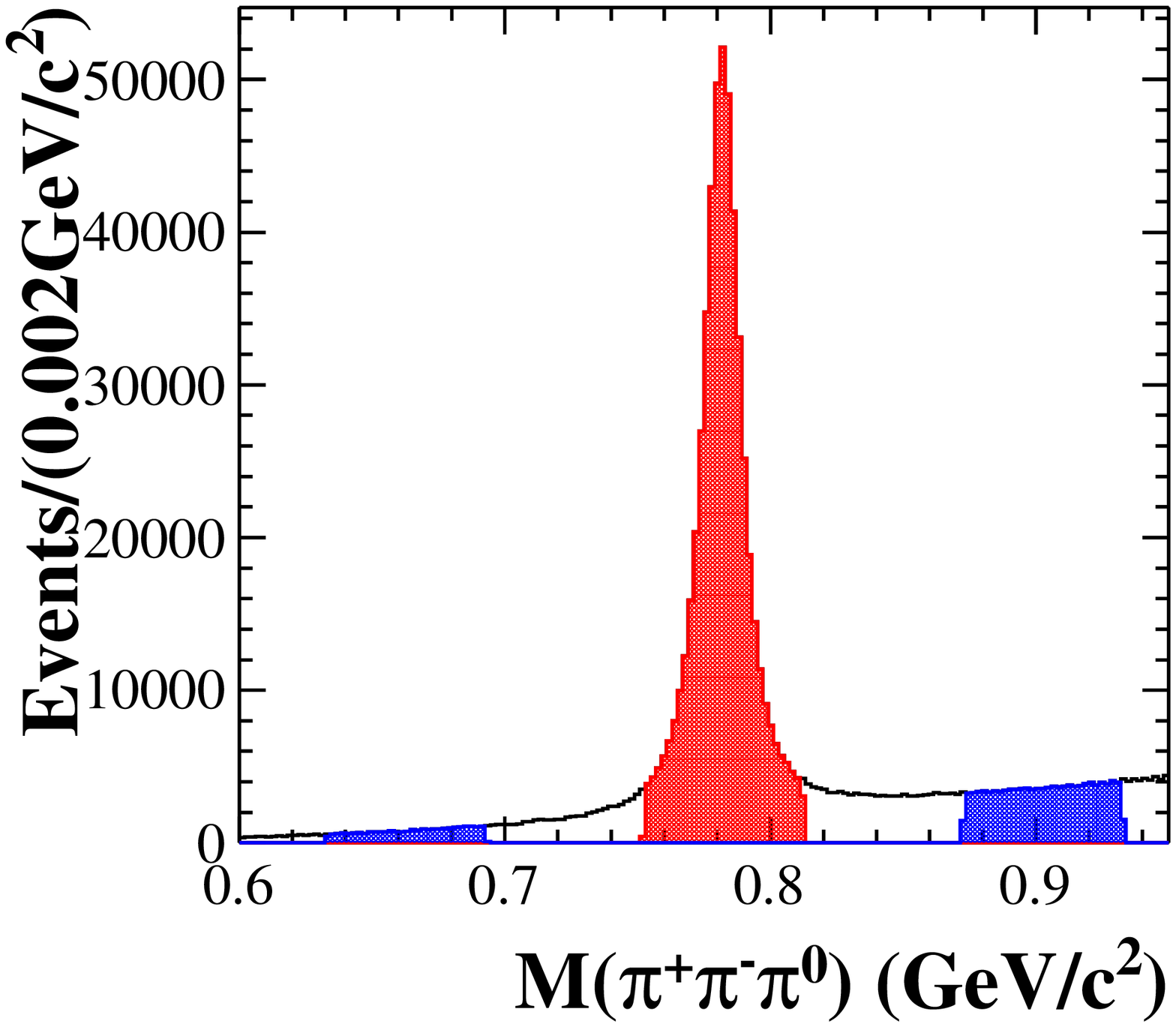}
\figcaption{\label{mw} The mass spectrum of $3\pi$. The middle hashed histogram shows the mass window of $\omega$ peak, and the left and right hashed histograms denote the sideband regions.}
\end{center}

\begin{center}
\includegraphics[width=3.8cm,height=3.5cm]{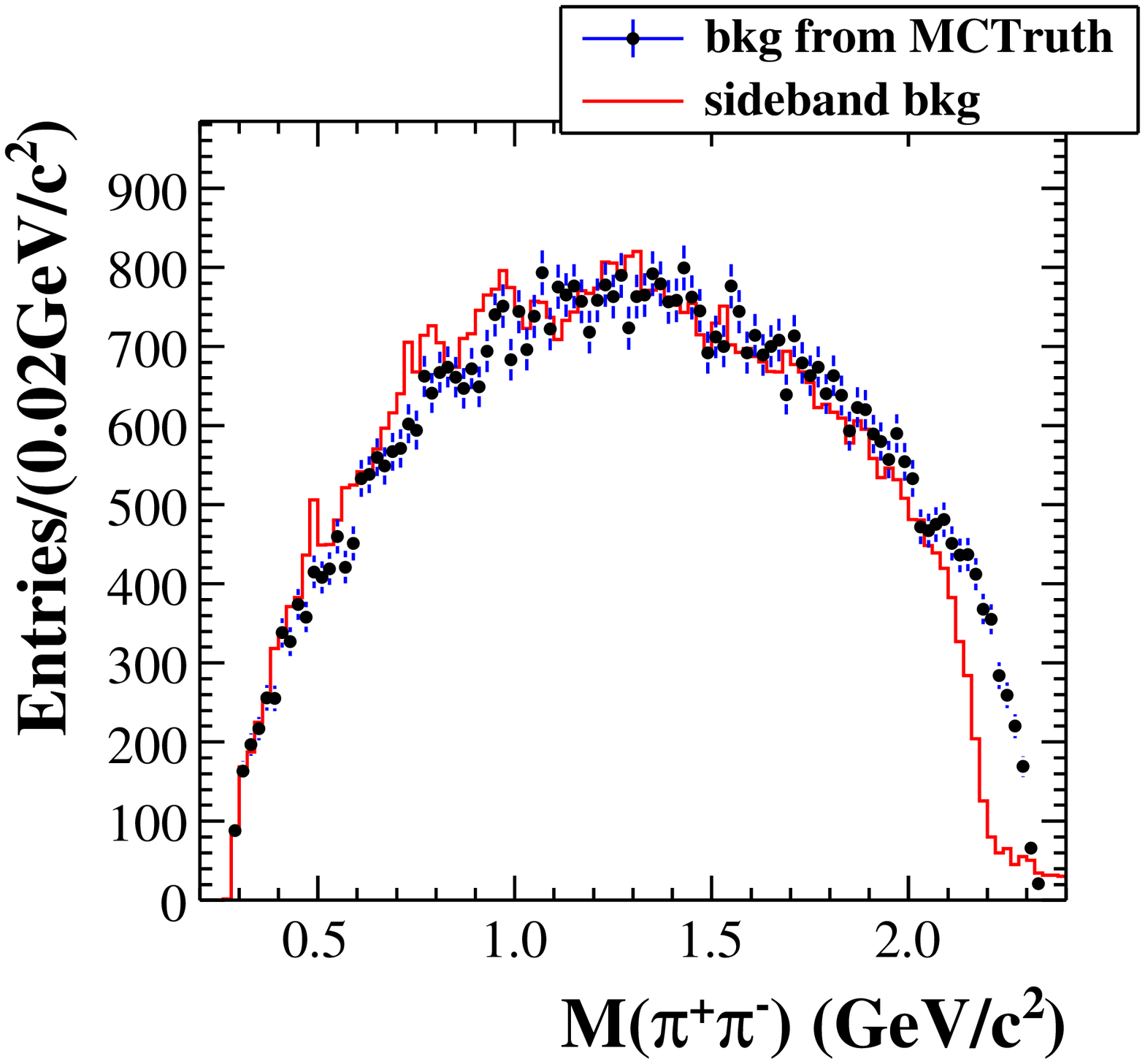}
\includegraphics[width=3.8cm,height=3.5cm]{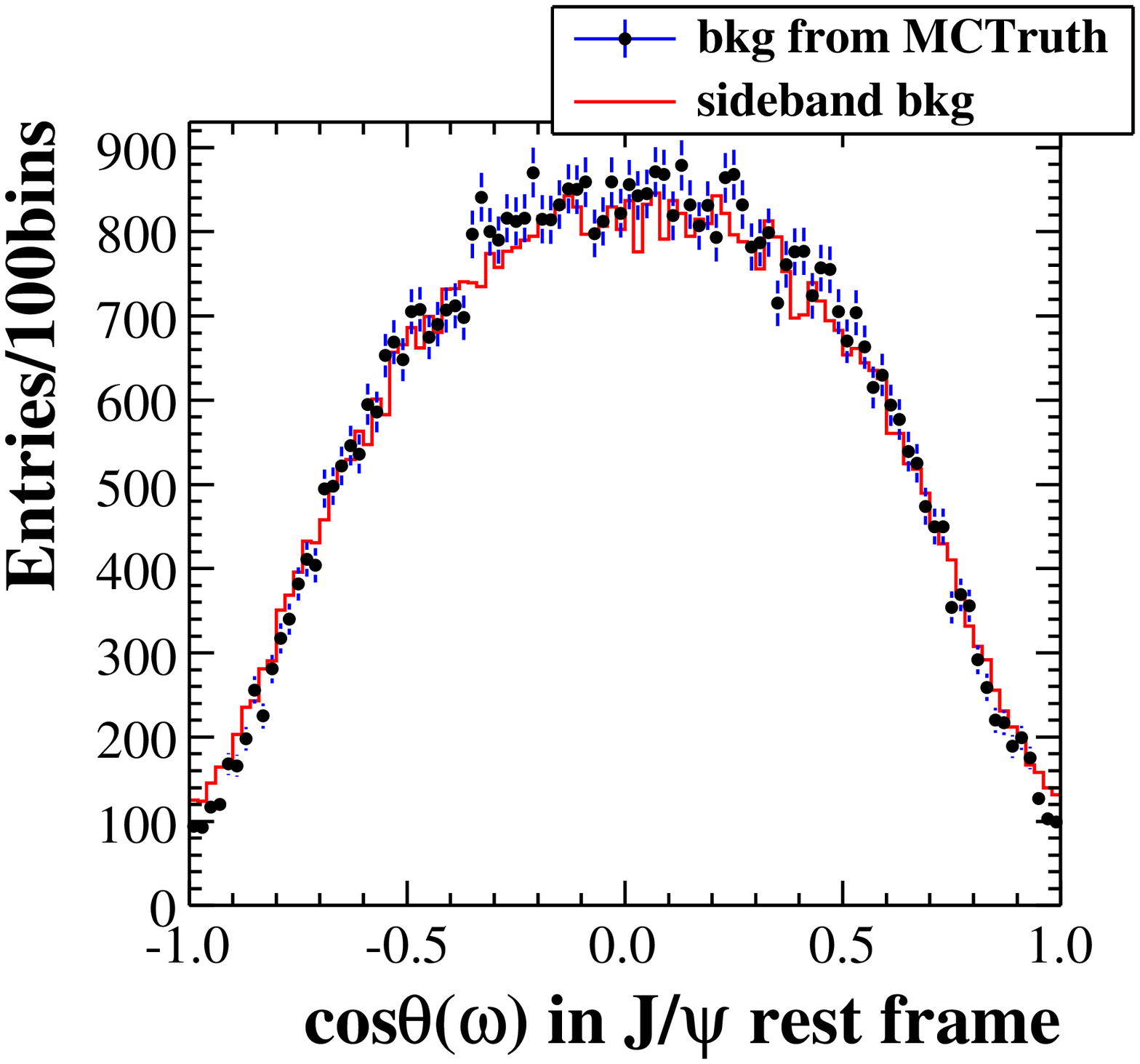}
\includegraphics[width=3.8cm,height=3.5cm]{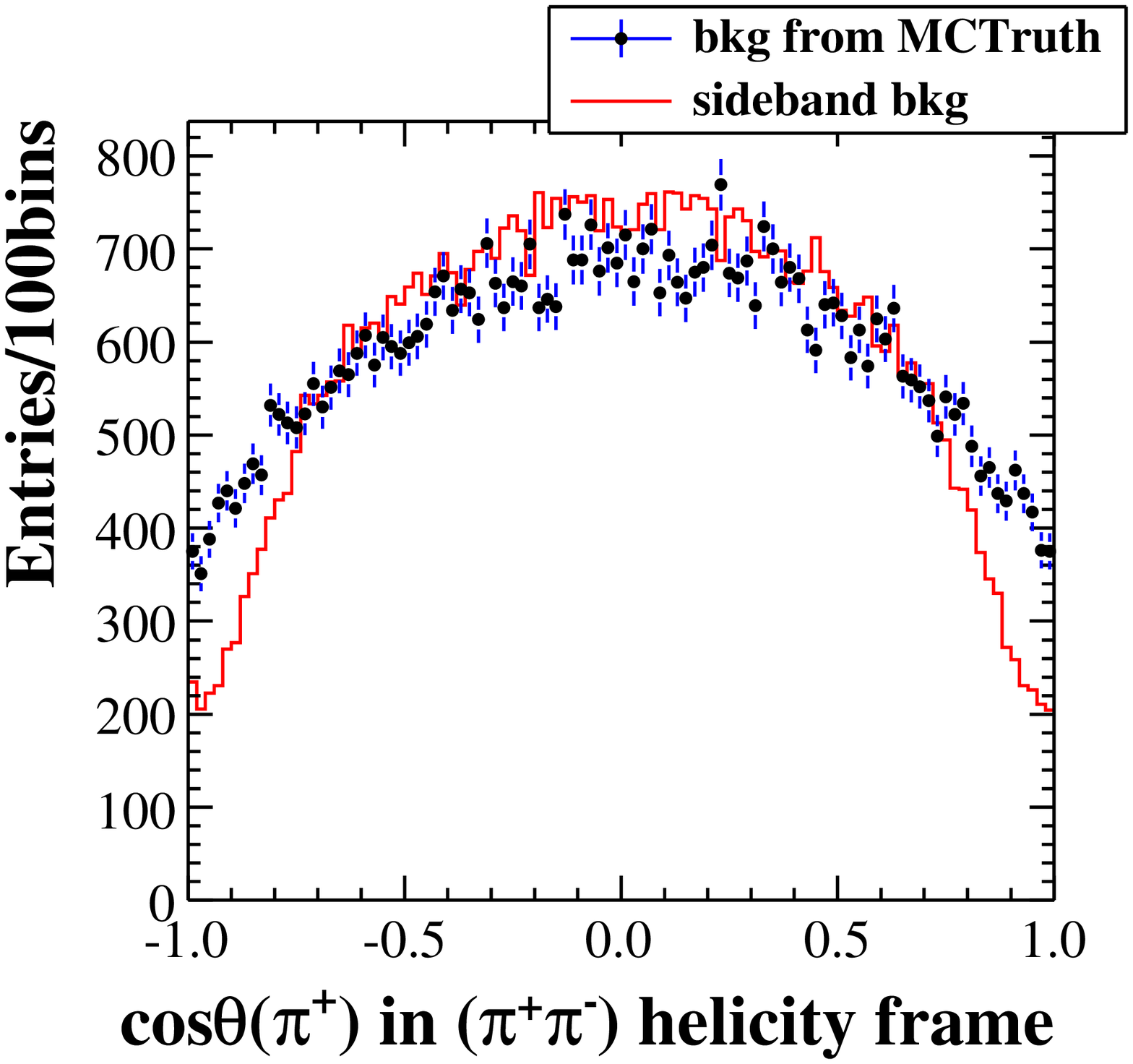}
\includegraphics[width=3.8cm,height=3.5cm]{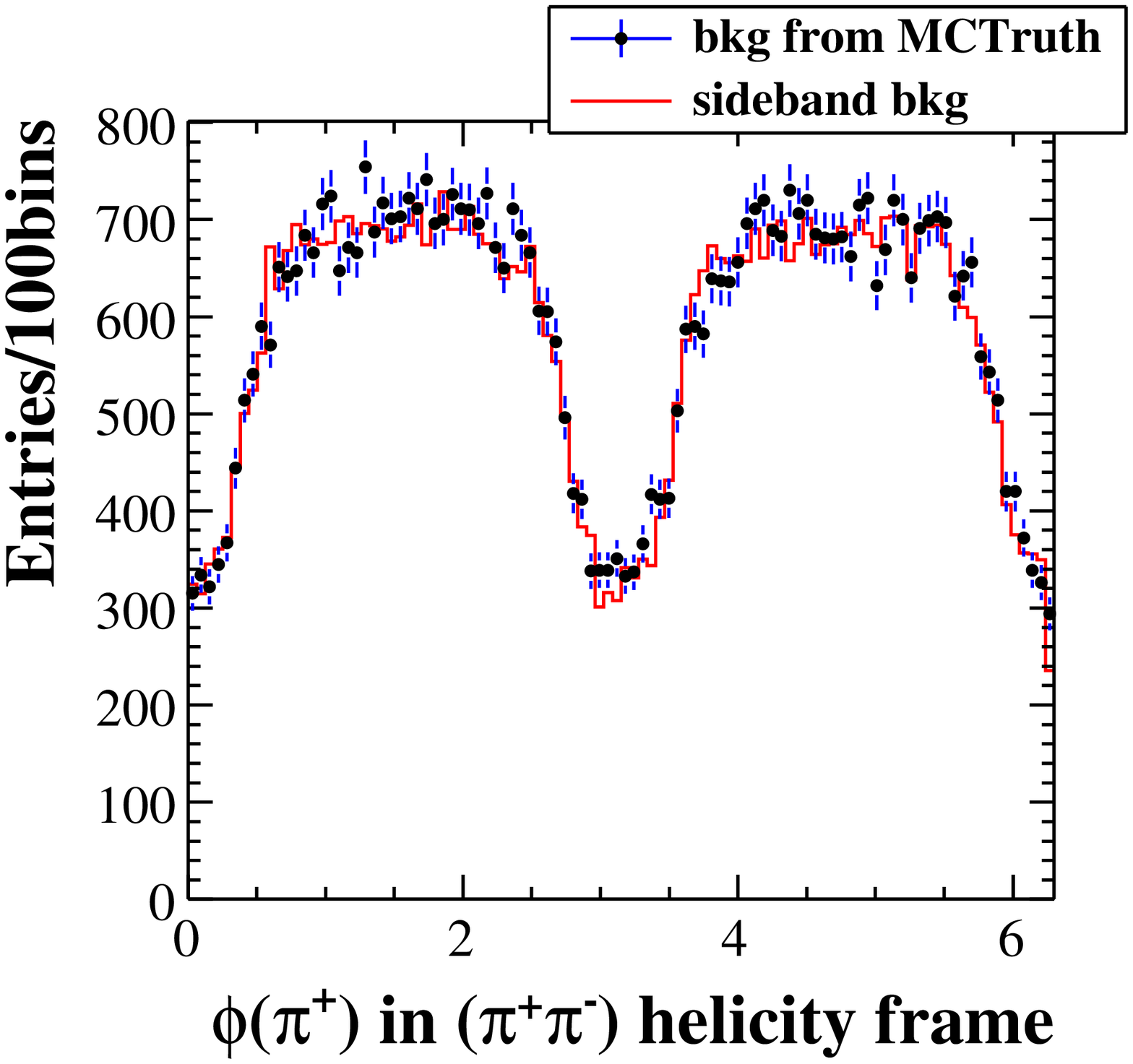}
\figcaption{\label{weight_DIY_wsb} Distributions of our analyzing variables. The dots with error bars show the true background distributions under $\omega$ peak, and the solid lines denote distributions estimated from $\omega$ sideband.}
\end{center}

\begin{center}
\tabcaption{ \label{tab_side} $\chi^{2}$/nbins of each variable comparison between backgrounds estimated by $\omega$ sideband and true distributions.}
\begin{tabular*}{80mm}{@{\extracolsep{\fill}}c|c|c|c|c}
\toprule
variable       & M($\pi^{+}\pi^{-}$)    & $\cos\theta(\omega)$   & $\cos\theta(\pi^{+})$    & $\phi(\pi^{+})$    \\ \hline
$\chi^{2}$     & \hphantom{00} 13.07    & \hphantom{0} 2.78      & \hphantom{0} 14.28       & \hphantom{0} 1.74     \\
\bottomrule
\end{tabular*}
\end{center}

\section{Probabilistic event weights}

Partial wave analysis is event based, so it would be advantageous if we could assign each
event a signal probability. This could then be used to weight the event's contribution to the likelihood
during fitting. It could also be used to weight the event's contribution to any distributions. In this way, we could subtract
the background without resorting to use events outside the signal region.

A method which involves using neighbor events to unfold the background from different sources to a data set has been developed in ref.~\cite{q-factor}. It manifests a concept of $_{s}Plot$ technique~\cite{splot} by generalizing the one-dimensional side-band subtraction method to higher dimensions. This method involves using nearest neighbor events to assign each event in a
data sample a quality factor (Q-factor) which gives a probability that it originates from a signal
sample. In ref.~\cite{q-factor}, the Q-factor is determined locally by fitting a control variable distribution of neighbor events. Eq.~\ref{eq_q_factor} gives the description of Q-factor, where $\vec{x}$ is a set of control variables, and $F_{s}(\vec{x}_{i})$ and $F_{b}(\vec{x}_{i})$ are the functional dependence for signal and background.
\begin{eqnarray}
Q_{i} = \frac {F_{s}(\vec{x}_{i})} {F_{s}(\vec{x}_{i}) + F_{b}(\vec{x}_{i})}
\label{eq_q_factor}
\end{eqnarray}

It is important to notice that, even though the control variable ($M(\pip\pim\piz)$ in our example) for background estimation is independent with the variable of interest (i.e. 4-momentum of $\omega\pip\pim$) used in analysis, the Q-factor of signal and background is dependent with the variable of interest, i.e. the location in the phase space. For instance, on the squared Dalitz plot of $\cos\theta(\pip)$ V.S. $M(\pip\pim)$, when $M(\pip\pim)\in(1.24,1.26)$ and $\cos\theta(\pip)\in(-0.96,-0.94)$, the signal-to-background ratio $S/B$ is $4.1$. When $M(\pip\pim)\in(1.74,1.76)$ and $\cos\theta(\pip)\in(-0.02,-0.00)$, the $S/B$ is $0.3$. $S$ and $B$ represent the event numbers of signal and background respectively. This is caused by different distributions of signal and background, which is shown in Fig.~\ref{sVb_DIY}. As pointed out by ref.~\cite{q-factor}, we have to perform a localized unfolding to obtain the Q-factor in different region of the phase space.

\begin{center}
\includegraphics[width=3.8cm,height=3.5cm]{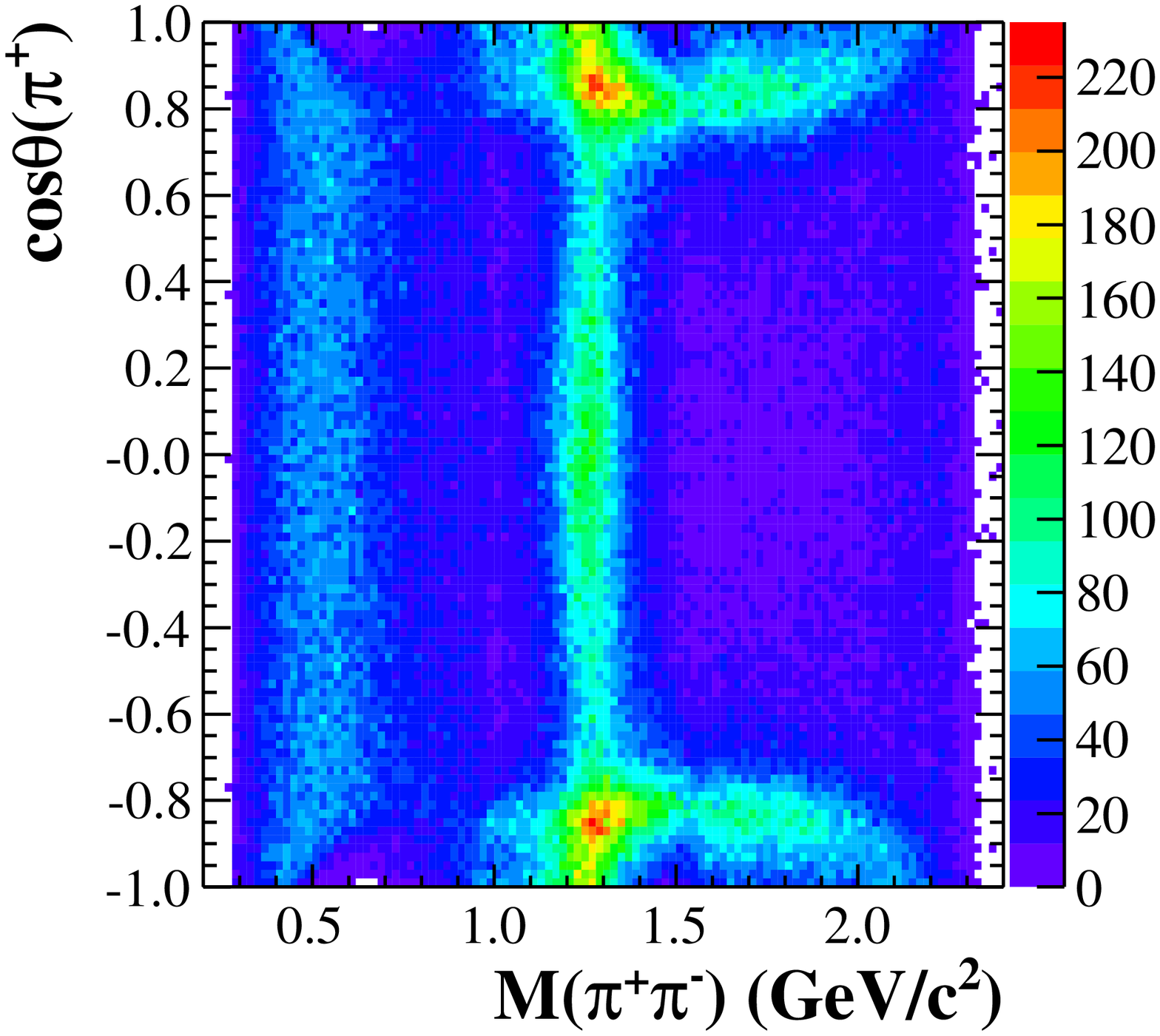}
\includegraphics[width=3.8cm,height=3.5cm]{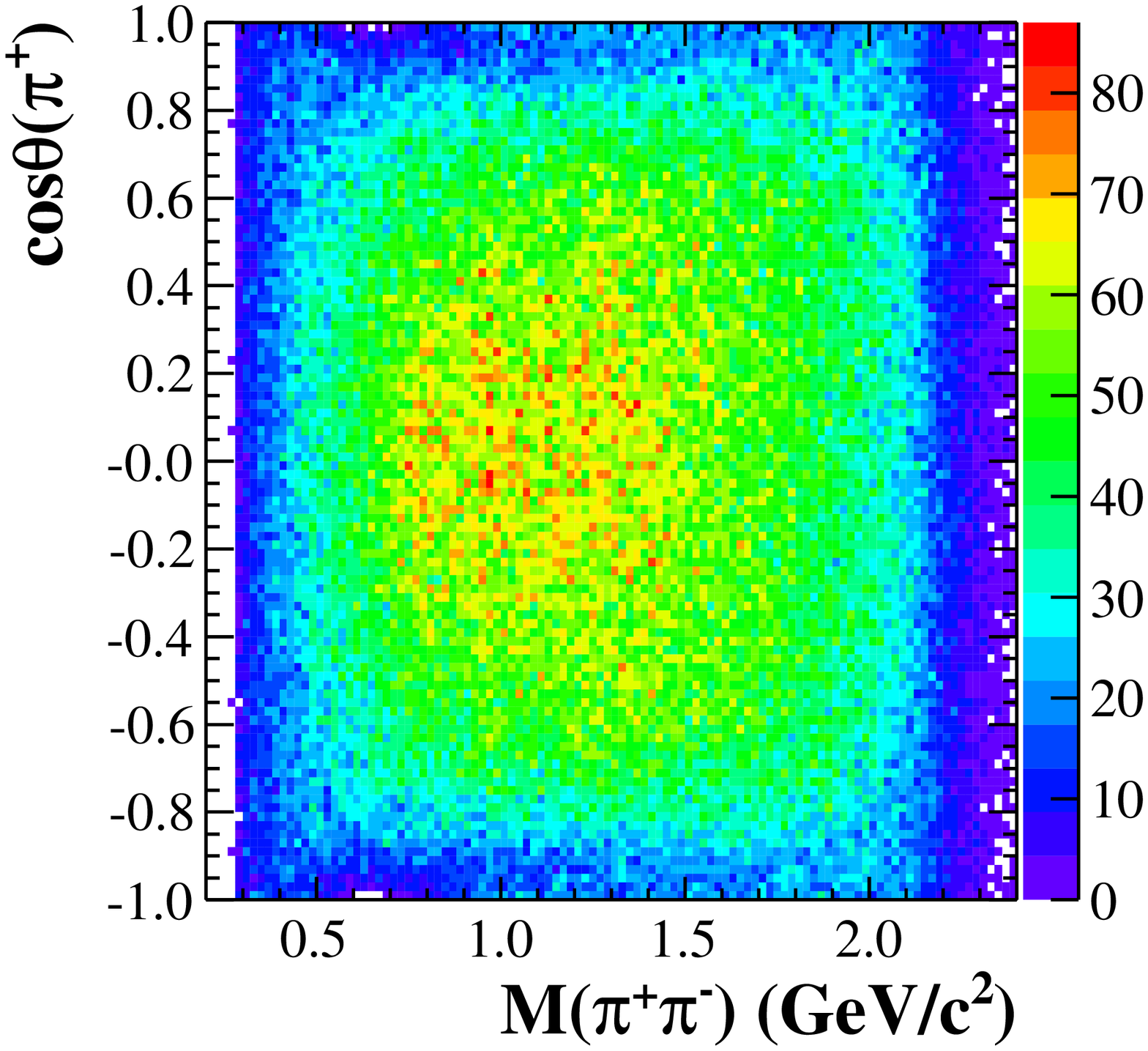}
\figcaption{\label{sVb_DIY} Dalitz plots from signal (left) and background (right) samples.}
\end{center}

Using the MC samples introduced in the previous section, one variable $x$, the invariant mass of $\pip\pim\piz$, is chosen as the control variable. We are interested in the 4-dimensional phase space of $J/\psi\to\omega\pip\pim$. For high statistics data sets, the practice of finding the neighborhoods event by event will be time consuming. In this test, we try to use an approximate approach. 
The Q-factor is determined locally in the squared Dalitz plot of $\jpsi\to\omega\pip\pim$: $\cos\theta(\pip)$ V.S. $M(\pip\pim)$  in $\pip\pim$ helicity frame. The distance to determine the neighborhood is then calculated on the squared Dalitz plot. For each bin of the squared Daltiz plot, we perform a fit to $M(\pip\pim\piz)$ with a Breit-Wigner (BW) convolved with a Gaussian resolution and a 2nd order polynomial $(PDF(x) = f\times BW(x)\bigotimes Gauss(x) + (1-f)\times poly(x))$. The Q-factor of an event in one bin is calculated with
\begin{eqnarray}
Q(x) = \frac{f\times BW(x)\bigotimes Gauss(x)}{f\times BW(x)\bigotimes Gauss(x) + (1-f)\times poly.(x)},\notag
\end{eqnarray}
where $f$ is the ratio between signal function and background function.
The variables of interest from backgrounds are shown in Fig.~\ref{weight_DIY_v11}. Table~\ref{tab_q_factor} lists the $\chi^{2}$/nbins of the comparisons between backgrounds estimated with probabilistic event weights and the real backgrounds.
\begin{center}
\includegraphics[width=3.8cm,height=3.5cm]{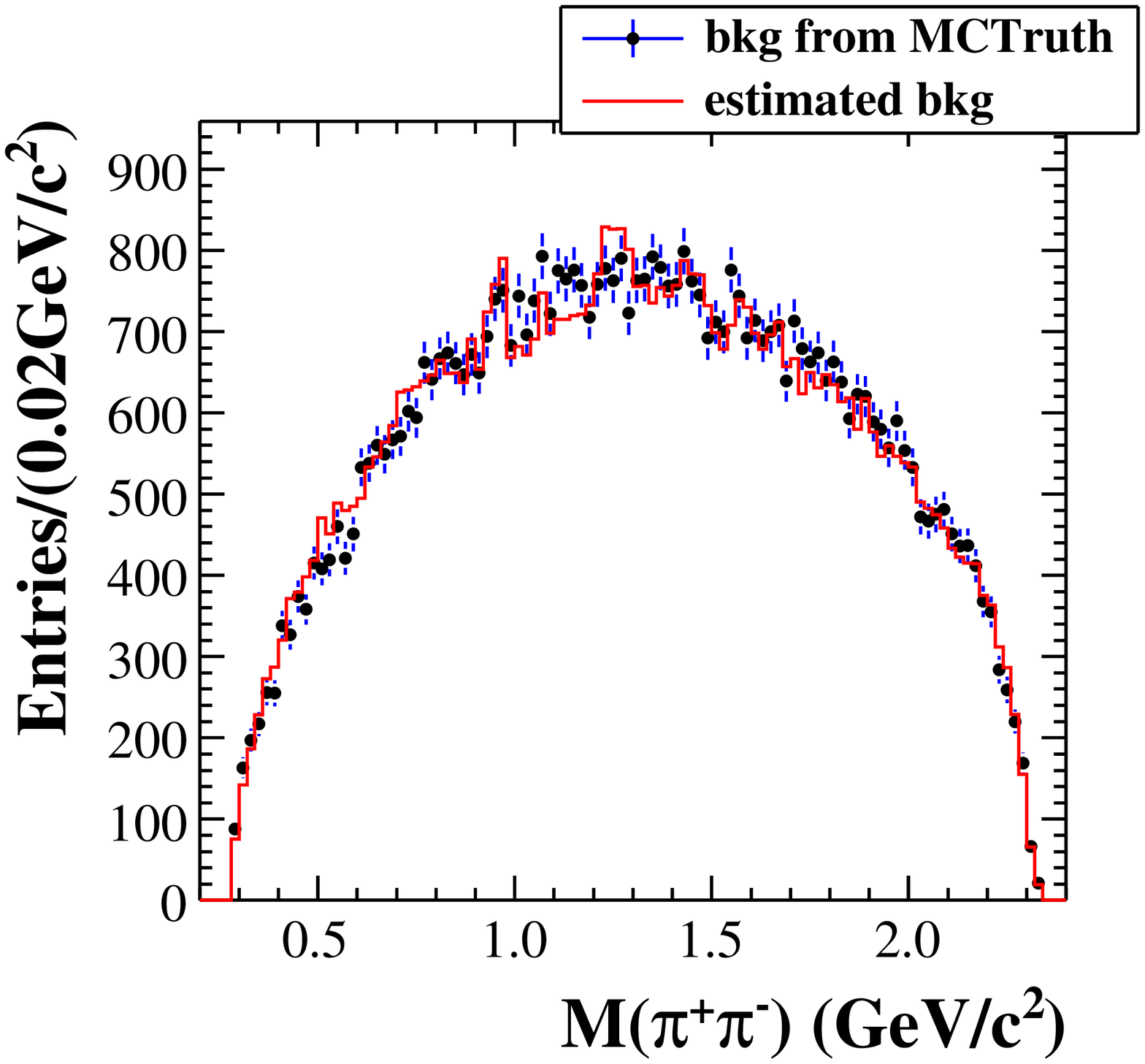}
\includegraphics[width=3.8cm,height=3.5cm]{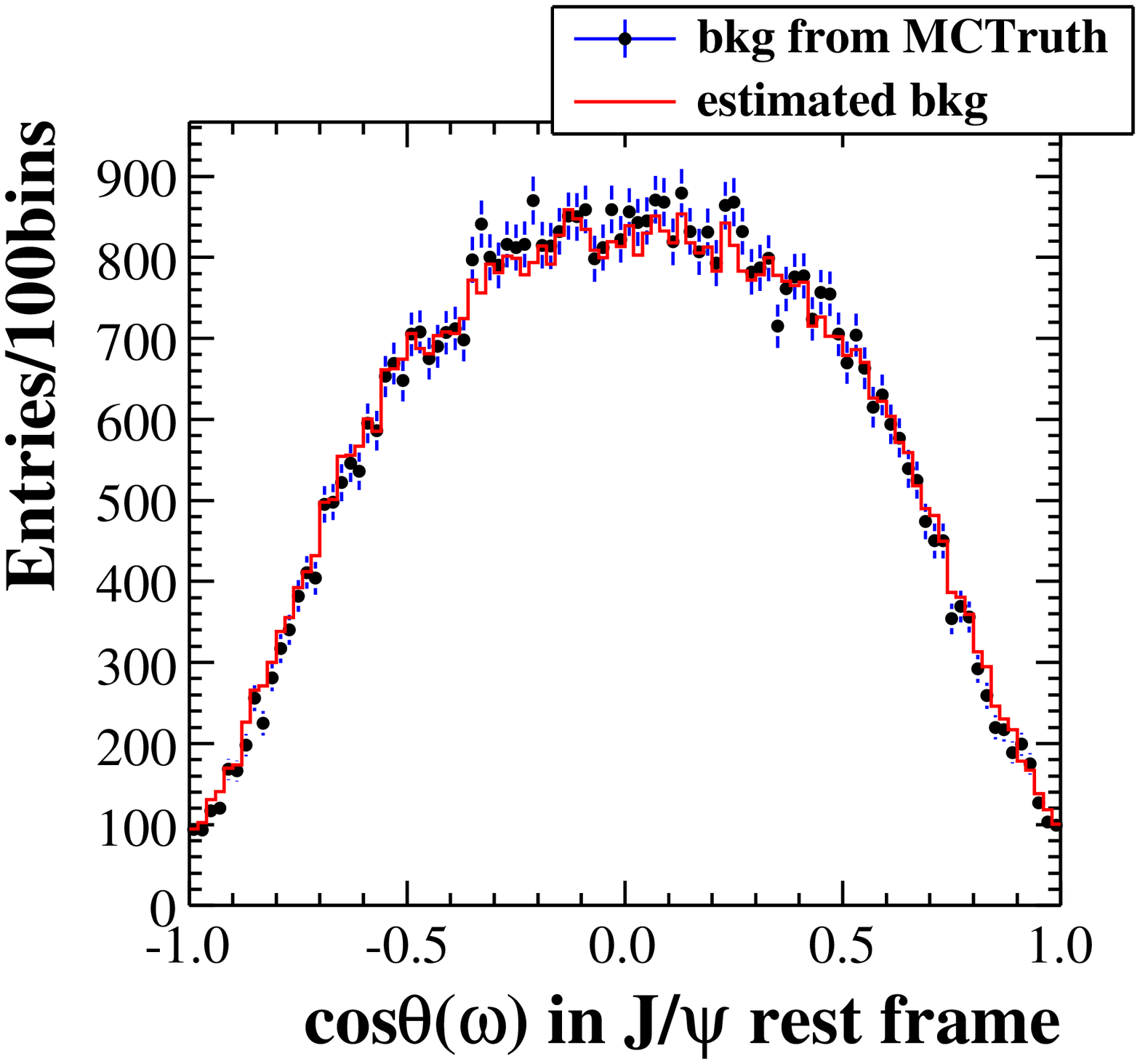}
\includegraphics[width=3.8cm,height=3.5cm]{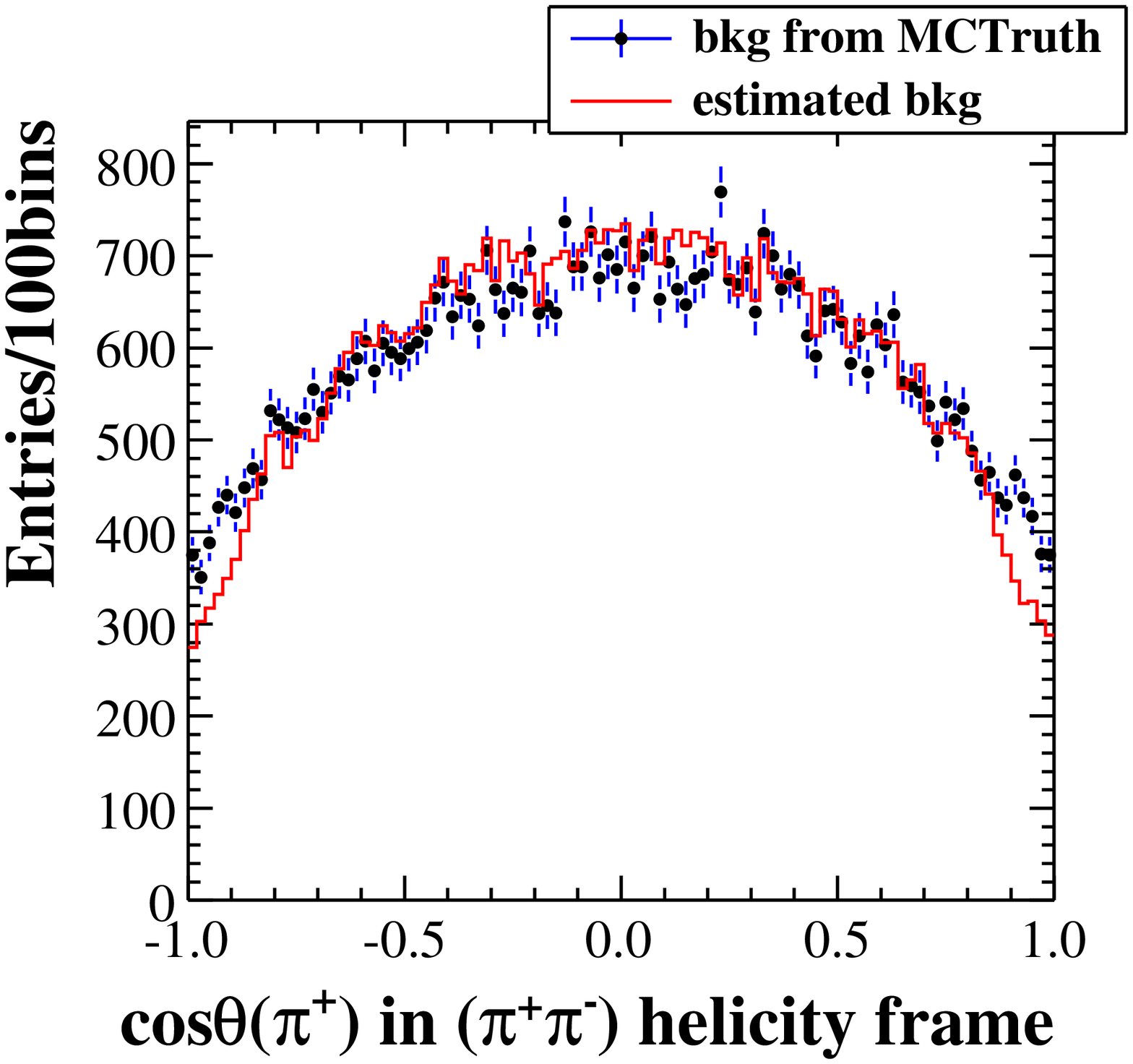}
\includegraphics[width=3.8cm,height=3.5cm]{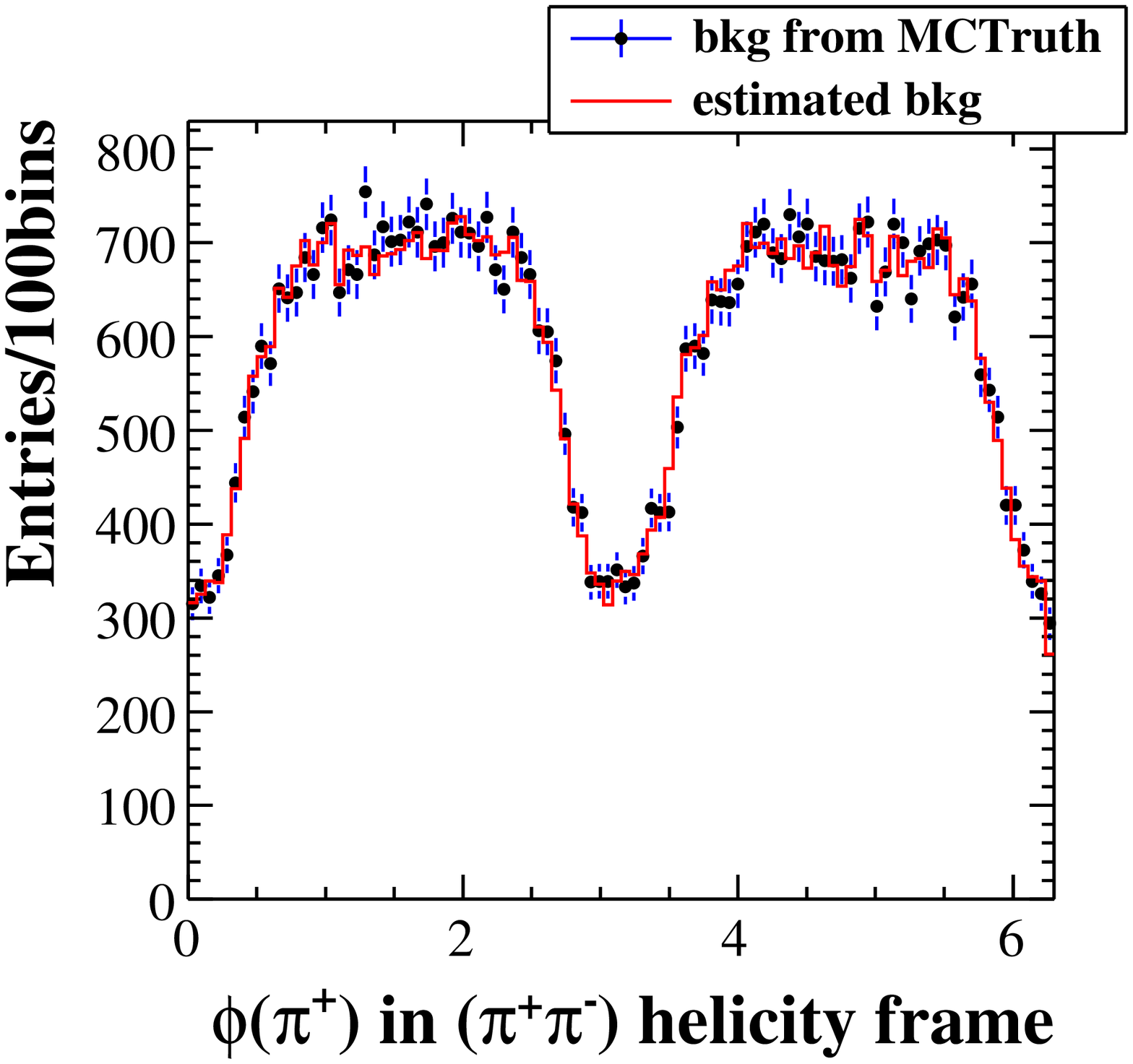}
\figcaption{\label{weight_DIY_v11} Distributions of our analyzing variables. The dots with error bars show the true background distributions under $\omega$ peak, and the solid histograms denote background distributions constructed with Q-factor.}
\end{center}

\begin{center}
\tabcaption{ \label{tab_q_factor} $\chi^{2}$/nbins of each variable comparison between backgrounds estimated with probabilistic event weights and true distributions.}
\begin{tabular*}{80mm}{@{\extracolsep{\fill}}c|c|c|c|c}
\toprule
variable          & M($\pi^{+}\pi^{-}$)  & $\cos\theta(\omega)$   & $\cos\theta(\pi^{+})$  & $\phi(\pi^{+})$  \\ \hline
$\chi^{2}$        & \hphantom{00} 1.50   & \hphantom{0} 1.07      & \hphantom{0} 3.34      & \hphantom{0} 0.78   \\
\bottomrule
\end{tabular*}
\end{center}

The consistence between the constructed background and the background from MC truth is obvious, and it is much better than that from the traditional sideband subtraction method. The test shows that the background subtraction method with probabilistic event weights is feasible in PWA at BES~III.  For example, with each event's probability, Eq.~\ref{eq_lnlikelyhood} in PWA could be rewritten as:
\begin{eqnarray}
\ln L = \sum\limits_{i=1}^{N} Q_{i} \times \ln \frac{\omega(\xi_{i})}{\int d\xi\omega(\xi_{i})\epsilon(\xi_{i})},
\end{eqnarray}
where the $Q_{i}$ is the signal probability of event $i$.


\section{Summary}

In this paper, the scenario has been pointed out that the sideband subtraction becomes problematic if the kinematics of the background are different from that of the signal. To extract physics from precise analyses, a novel background subtraction method with probabilistic event weights has been introduced. The feasibility has been studied with MC samples. The results show that, for a 3-body $\jpsi$ decay, the backgrounds could be estimated with the Q-factors determined in Dalitz plot bins.  When applying the procedure in a specific analysis, dedicated studies on the choice of control variables, the choice of neighborhoods and the determination of Q-factors are suggested.


\vspace{10mm}
\acknowledgments{The authors are very grateful to the BESIII-PWA group for their constructive suggestions. And the authors are also very grateful to the USTC-BES working group for the useful discussions.}

\end{multicols}

\vspace{10mm}

\vspace{-1mm}
\centerline{\rule{80mm}{0.1pt}}
\vspace{2mm}

\begin{multicols}{2}

\end{multicols}

\clearpage

\end{document}